\documentclass[pra,aps,twocolumn,superscriptaddress,amsmath,amssymb,nofootinbib]{revtex4-1}
\usepackage{graphicx}
\usepackage{epstopdf}
\usepackage{units}
\usepackage{color}

\newcommand{\bra}[1]{\langle#1|}
\newcommand{\ket}[1]{|#1\rangle}

\begin{document}

\title{Localization transition in presence of cavity backaction}
\date{\today}

\author{Katharina Rojan}
\affiliation{Theoretische Physik, Universit\"at des Saarlandes, D-66123 Saarbr\"ucken, Germany}
\affiliation{Universit\'e Grenoble-Alpes, LPMMC, BP166, F-38042 Grenoble, France}
\affiliation{CNRS, LPMMC, BP166, F-38042 Grenoble, France}
\author{Rebecca Kraus}
\affiliation{Theoretische Physik, Universit\"at des Saarlandes, 
  D-66123 Saarbr\"ucken, Germany}
\author{Thom\'as Fogarty}
\affiliation{Theoretische Physik, Universit\"at des Saarlandes, 
  D-66123 Saarbr\"ucken, Germany}
  \affiliation{Quantum Systems Unit, Okinawa Institute of Science and Technology Graduate
University, Okinawa, 904-0495, Japan}
\author{Hessam Habibian}
\altaffiliation{currently at Accenture S.L.U., 08174 Sant Cugat del Vall\'es, Barcelona, Spain}
\affiliation{Theoretische Physik, Universit\"at des Saarlandes, 
  D-66123 Saarbr\"ucken, Germany}
  \author{Anna Minguzzi}
\affiliation{Universit\'e Grenoble-Alpes, LPMMC, BP166, F-38042 Grenoble, France}
\affiliation{CNRS, LPMMC, BP166, F-38042 Grenoble, France}
\author{Giovanna Morigi}
\affiliation{Theoretische Physik, Universit\"at des Saarlandes, 
  D-66123 Saarbr\"ucken, Germany} 

\begin{abstract}
We study the localization transition of an atom confined by an external optical lattice in a high-finesse cavity. The atom-cavity coupling yields an effective secondary lattice potential, whose wavelength is incommensurate with the periodicity of the optical lattice. The cavity lattice can induce localization of the atomic wave function analogously to the Aubry-Andr\'e localization transition. Starting from the master equation for the cavity and the atom we perform a mapping of the system dynamics to a Hubbard Hamiltonian, which can be reduced to the Harper's Hamiltonian in appropriate limits. We evaluate the phase diagram for the atom ground state and show that the transition between extended and localized wavefunction is controlled by the strength of the cavity nonlinearity, which determines the size of the localized region and the behaviour of the Lyapunov exponent. The Lyapunov exponent, in particular, exhibits resonance-like behaviour in correspondence with the optomechanical resonances. Finally we discuss the experimental feasibility of these predictions.
\end{abstract}

\maketitle

\section{Introduction}

Cavity quantum electrodynamics (CQED) with cold atoms provides a rich framework to study the wave-particle duality of light and matter \cite{Walther:2006,Kimble:2005,Ritsch:2013}. In this environment, the interaction of a single photon with a single atom has been brought to a level of control that is sensitive to the finite spatial localization of the atom within the cavity mode \cite{Hood:2000,Pinkse:2000,Guthoerlein:2001,Casabone:2013,Reiserer:2014,Rempe:2015}. This property is at the basis of several protocols, which exploit the optomechanical coupling between atoms and photons in CQED in order to cool the atomic motion \cite{Ritsch:2013,Horak:1997,Vuletic:2000,Pinkse:2004,Kampschulte:2014}, to perform high precision measurements \cite{Stamper-Kurn:2015,Haas:2014}, and to create novel sources of quantum light \cite{Parkins:1999,Morigi:2006,Reimann:2014,Ritter:2016}, to provide some examples.

Cavity backaction, moreover, modifies the dynamics to the extent that photons and atoms become strongly correlated: Since the photon field depends on the atomic position within the resonator, the mechanical forces that the atom (or molecule) experiences depend on the center-of-mass wave function within the cavity mode \cite{Domokos:2003,Schuetz:2013}. This nonlinearity is at the basis of several collective phenomena, such as the formation of spatial patterns \cite{Black:2003,Baumann:2010,Baumann:2011}, synchronization \cite{Cube:2004,Zhu:2015}, and exotic phases of ultracold matter \cite{Larson:2008,Wolke:2012,Klinder:2015,Donner:2015}. Even at the level of a single particle it can give rise to peculiar behaviours, which shed light on the interplay between nonlinear dynamics and quantum fluctuations. 

\begin{figure}[htbp]
\begin{center}
  (a)\vtop{\vskip-0ex\hbox{\includegraphics[width=0.4\textwidth]{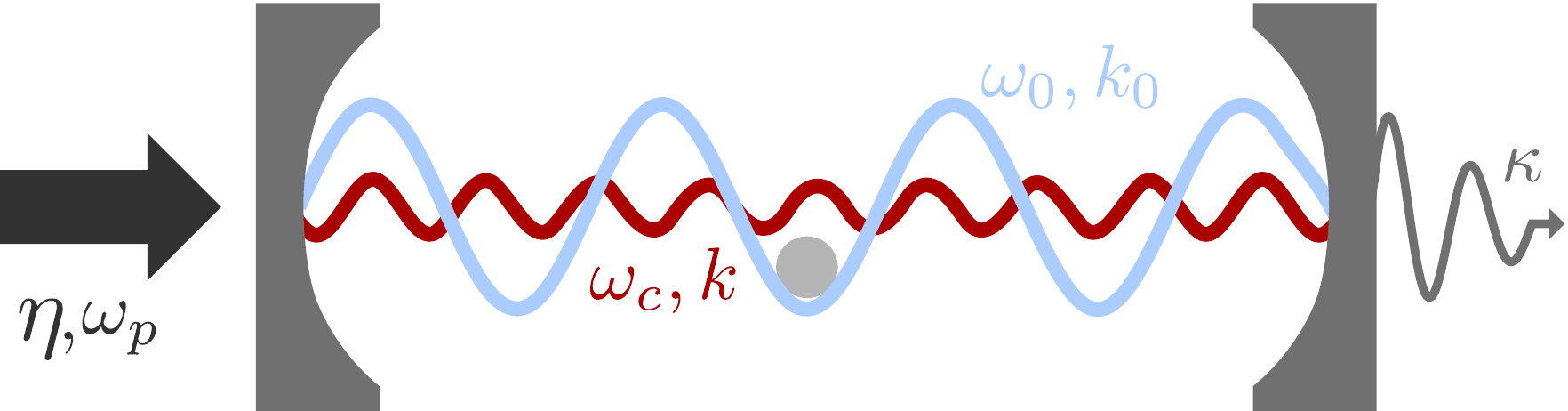}}}\\
  (b)\vtop{\vskip-0ex\hbox{\includegraphics[width=0.4\textwidth]{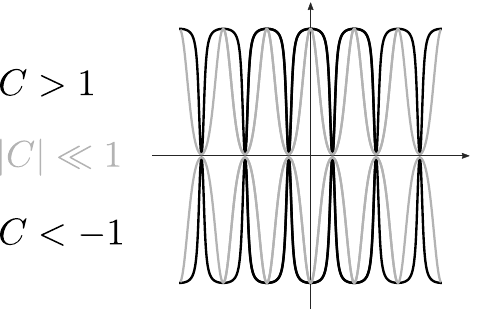}}}
\end{center}
\caption{\label{Fig1}(Color online)(a) A single atom is confined by an optical lattice with wave number $k_0=2\pi/\lambda_0$ within a standing-wave resonator. Its motion optomechanically interacts with a high-finesse mode at frequency $\omega_c$ and wave number $k=2\pi/\lambda$, whose wavelength $\lambda$ is incommensurate with the optical lattice periodicity $\lambda_0/2$. The depth of the cavity lattice is determined by the balance between a pump, with strength $\eta$ and frequency $\omega_p$, and the losses at rate $\kappa$. We study the localization transition in this setup, where the nonlinearity due to strong coupling with the cavity (given by the cooperativity $C$) modifies the effective incommensurate potential. The optomechanical potential is illustrated in (b)  as a function of $x$ for three different values of the cooperativity $C$ and when the laser is resonant with the cavity. The limit $|C|\ll1$ corresponds to the Aubry-Andr\'e model. }
\end{figure}

In this work, we theoretically investigate the regime in which cavity backaction can induce the transition to localization of the atomic center-of-mass wave function. The system we consider is illustrated in Fig.\ref{Fig1}(a): a single atom is tightly confined by an external optical lattice within a high-finesse cavity, its dipole strongly couples with a standing-wave mode of the resonator. In the regime in which this coupling is dispersive, the mechanical effects of the cavity field are described by a second periodic potential. We choose the two lattice wavelengths with periods which are incommensurate with each other. The combination of these two characteristic lengths gives rise to an aperiodic, pseudorandom potential. 

In the limit where the cavity nonlinearity can be neglected, the system is described by the Aubry-Andr\'e model \cite{Aubry} or the Harper model \cite{Harper}, which predicts a transition from an extended to a localized phase when the ratio between the depths of the two potentials exceeds a critical value. This localization transition has been observed experimentally with ultracold atoms confined by bichromatic optical lattices by tuning the amplitude of the secondary lattice potential \cite{Roati2008,Schreiber2015,Derrico2013}. The effect of interactions on the Aubry-Andr\'e model  has been investigated theoretically both in the mean-field weakly interacting regime \cite{Modugno2010} as well as for arbitrary interactions at low lattice filling \cite{Xiao2008,Roux2008}. Quasiperiodic potentials have also been realised with exciton polaritons in semiconductor microcavities \cite{Tanese2014}.

Differing from these realisations, the strong coupling with the cavity introduces a novel feature: The depth of the cavity potential, in fact, is proportional to the number of intracavity photons, which is a dynamical variable coupling optomechanically with the atomic motion. In this setting we analyze the localization transition and discuss possible experimental regimes where it could be observed. 

This manuscript is organized as follows. In Sec. \ref{Sec:Model} we introduce the theoretical model, which encompasses the effect of the cavity nonlinearity, and show how it results from the optomechanical coupling of a single atom with the single mode of a lossy cavity. In Sec. \ref{Sec:PD} we analyze the phase diagram for the ground state as a function of the cavity parameters and then discuss experimental realizations in cavity QED setups. In Sec. \ref{Sec:Conclusions} we draw the conclusions and present outlooks to this work. 

\section{Aubry-Andr\'e's model in presence of cavity backaction}
\label{Sec:Model}

In this section we discuss the theoretical model at the basis of our analysis, which is a Hubbard model with the onsite energy resulting from a second, incommensurate potential. We then identify the parameters for which one recovers the Aubry-Andr\'e \cite{Aubry} or Harper model \cite{Harper}. We further discuss the conditions under which the Hubbard model describes a cold atom which opto-mechanically interacts with the mode of a high-finesse optical cavity.

\subsection{Hubbard model for cavity QED with cold atoms}

The model at the basis of our analysis results from the one dimensional dynamics of a particle of mass $m$ in two periodic potentials, of which one, denoted by $\hat W_{\rm{ext}}(\hat x)$, tightly traps the particle at its minima while the second, $\hat V_{\rm{eff}}(\hat x)$, is a perturbation to the first potential, $\hat H_{\rm{eff}}=\hat p^2/(2m)+\hat W_{\rm{ext}}(\hat x)+\hat V_{\rm{eff}}(\hat x)$, with $\hat p$ and $\hat x$ the canonically-conjugate momentum and position. The cavity and external potentials are periodic with wave numbers $k$ and $k_0$, respectively, where $k=\beta k_0$ and $\beta$ is an irrational number.  Therefore, the Hamiltonian is aperiodic. Specifically $\hat W_{\rm{ext}}(\hat x)=\hat W_{\rm{ext}}(\hat x+\pi/k_0)$, while $\hat V_{\rm{eff}}(\hat x)=\hat V_{\rm{eff}}(\hat x+\pi/k)$. For later convenience we write $\hat V_{\rm{eff}}(\hat x)=v_0f(\hat x)$, where $v_0$  has the dimensions of an energy, and $f(\hat x)=f(\hat x+\pi/k)$ is a dimensionless function. In the limit in which the dynamics can be restricted to the lowest band of the deep lattice $\hat W_{\rm{ext}}(\hat x)$ \cite{Guarrera2007}, we can describe it by means of the Hubbard Hamiltonian
\begin{align}\label{Hfinal}
 \hat H_{BH}= -t\sum\limits_{n=1}^{L-1}\left(\ket{n}\bra{n+1}+\ket{n+1}\bra{n}\right)+\sum\limits_{n=1}^{L}\delta \epsilon_{n}\ket{n}\bra{n}\,,
\end{align}
where $\ket{n}$ denotes the state of the particle at site $n$ of the external lattice potential $\hat W_{\rm{ext}}$, with $n=1,\ldots,L$ and $L$ the total number of sites. The Hubbard Hamiltonian is composed of the hopping term, scaled by the tunneling coefficient $t=\bra{n}\hat p^2/(2m)+\hat W_{\rm{ext}}(\hat x)\ket{n+1}$, and by the diagonal term in the basis $\{\ket{n}\}$, whose coefficients are the onsite energy $\delta \epsilon_{n}=\bra{n}\hat{H}_{eff}\ket{n}$ and which are site dependent since the Hamiltonian is aperiodic. After subtracting an arbitrary energy constant, we can rewrite these coefficients as 
\begin{equation}\label{Harper}
  \delta\epsilon_n=\bra{n}\hat V_{\rm{eff}}(\hat x)\ket{n}=v_0
  \int_{-L_0/2}^{L_0/2} dx \ \ w_n(x) f(x)w_n(x)\,,
\end{equation}
where $L_0=L\pi /k_0$ and $w_n(x)=\langle x\ket{n}$ are the Wannier functions, which are real valued \cite{Kohn}.

In the Aubry-Andr\'e model the site-dependent onsite potential $\delta\epsilon_n$ has the form
\begin{align}\label{onsite_AA}
\delta\epsilon_n=v_0\cos(2\pi\beta n). 
\end{align}
The self-duality of the model \cite{Aubry} allows to show that a continuous transition occurs for the ground state when the value of the energy $v_0$ reaches a critical potential strength $v_c^{AA}=2t$ \cite{Harper,WobstNJP}. If $v_0<v_c^{AA}$ the ground state wavefunction is spatially extended, while for $v_0>v_c^{AA}$ the wavefunction decays exponentially indicating Anderson-like localization \cite{Anderson1956,Aubry,WobstNJP}.

In this work, we analyze the localization transition when the incommensurate potential is given by the function
\begin{equation}
\label{atan}
f(x)=\arctan\left(-\delta_c'+C\cos^2(\beta k_0x)\right)\,,
\end{equation}
and thus contains several harmonics of the Aubry-Andr\'e potential, given in equation \eqref{onsite_AA}. This functional form is reminiscent of the one considered in Ref. \cite{Peyrard:1982} and is typically encountered in optomechanical problems in CQED \cite{Asboth:2004,Larson:2008,Cormick:2013,Fogarty:2015}, as we detail in Sec. \ref{Sec:CQED}. The parameters $\delta_c'$ and $C$ are real valued and can take both positive and negative values. The parameter $\delta_c'$ is responsible for the appearance of non-trivial poles which can depend on the form of the ground state wavefunction and on $C$. The parameter $C$ controls the functional form of the cavity-induced potential $f(x)$, as illustrated in Fig. \ref{Fig1}(b). For $|C|\ll 1$ the onsite energy essentially reduces to Eq. \eqref{Harper} with $f(x)=\cos(2\beta k_0 x)$, however with the new amplitude $v_0'=|C|v_0/[2(\delta_c'^2+1)]$. In this limit, we have demonstrated \cite{Rojan:2016} that the critical value at which localization occurs is found at $v_c'=2t/\alpha$, giving
\begin{equation}
\label{v0:c}
v_c^{cav}=\frac{4t}{\alpha}\frac{\delta_c'^2+1}{|C|}\,,
\end{equation}
where the correcting factor $\alpha$ reads $\alpha=\sqrt{A^2+B^2}$, with \mbox{$A=-\int dx\ \ w_0^2(x) \sin(2\beta k_0 x)$} and \mbox{$B=\int dx \ \ w_0^2(x) \cos(2\beta k_0 x)$} \cite{Rojan:2016}.
The parameter $|C|$ corresponds to the cooperativity of CQED, which measures the strength of cavity backaction on the atom's scattering properties \cite{Kimble:1994}. The strong-coupling regime in CQED is typically characterized by $|C|\simeq 1$ or larger, which is the regime where higher harmonics of the optomechanical potential become relevant. Differing from the Aubry-Andr\'e model, this is the regime where the model is not self-dual. 

In this paper we study the transition to spatial localization. We determine numerically the ground state $\ket{\Psi_0}$ of Hamiltonian \eqref{Hfinal} with potential \eqref{atan} as a function of $C$, $\delta_c'$, and the ratio $v_0/t$. We characterize the transition by means of the inverse participation ratio $P_x$ \cite{Thouless:1974}
\begin{equation}\label{IPR}
P_x = \sum\limits_{n=1}^{L}|\langle n|\Psi_0\rangle|^4\,.
\end{equation}
The inverse participation ratio (IPR) is of the order $1/L$ if the atom spatial wavefunction is uniform over the lattice, whereas it approaches unity when the atom is localized on one single lattice site. 

We also monitor the degree of localization by the Lyapunov exponent,  defined as \cite{Aubry}
\begin{equation}\label{Lyapunov}
 \gamma=-\lim_{n\rightarrow\infty}\frac{\log(|\langle n|\Psi_0\rangle|^2)}{2n}\,.
 \end{equation}
According to Thouless' formula \cite{Thouless:1983}, in the localized regime of the Aubry-Andr\'e model it reads 
\begin{equation}\label{crit_exp}
 \gamma=\log\left(\frac{v_0}{v_c}\right)\,,
\end{equation}
with $v_c=v_c^{\rm{AA}}$ in the case of the original Aubry-Andr\'e model, i.e. $\delta\epsilon_n$ is given by equation \eqref{onsite_AA}, and $v_c=v_c^{\rm{cav}}$ Eq.~\eqref{v0:c}. 
In our calculation we obtain the Lyapunov exponent $\gamma$ by fitting the spatial decay of the wavefunction by means of an exponential function.

\subsection{Self-induced localization in cavity QED}
\label{Sec:CQED}

In this section we derive the the Hubbard Hamiltonian of Eq. \eqref{Hfinal} starting from the master equation for the density matrix $\rho$ of a linearly polarizable particle and of a lossy cavity field, which strongly couple to one another by means of an optomechanical interaction \cite{Domokos}. In the case of an atom, its dipolar transition is assumed to be driven far-off resonance by the fields, so that spontaneous decay can be neglected within the typical time scales we consider. 

\subsubsection{Master equation}

The relevant degrees of freedom for the atom are the momentum $\hat p$ along $x$ and the canonically-conjugated position $\hat x$, and the cavity mode degrees of freedom are the photon annihilation and creation operators $ \hat a$ and $\hat a^{\dagger}$, respectively, with the commutation relation $[\hat a, \hat a^{\dagger}]=\hat 1$. We denote by $m$ the atomic mass and by $\omega_c$ the cavity mode frequency, with wavelength $\lambda=2\pi c/\omega_c$, wavenumber $k=\beta k_0$ and spatial mode function $\cos(\beta k_0x)$. 

The system is driven by a laser, which is described by a classical field. The laser frequency $\omega_p$ is the reference frequency: the atom transition frequency $\omega_0$ is given by the detuning $\Delta_a=\omega_p-\omega_0$ and the cavity mode frequency by the detuning $\delta_c=\omega_p-\omega_c$. In the limit in which $|\Delta_a|$ is the largest frequency characterizing the dynamics, the atom's internal degrees of freedom are eliminated in second-order perturbation theory:  In this regime the atomic dipole behaves as a classical dipole, and its response to the field is described by its polarizability.  The dynamics of the density matrix $\rho(t)$ describing the state of the atomic center-of-mass position and of the cavity field is governed by the master equation  
\begin{equation}
\partial_t\hat\rho=\frac{1}{{\rm i}\hbar}[\hat H,\hat \rho]+\mathcal L \hat \rho\,,
\end{equation}
where Hamiltonian $\hat H$ describes the coherent optomechanical dynamics coupling between the atom's motion and the cavity mode, and the dissipator $\mathcal L$ describes cavity losses at rate $\kappa$:
\begin{equation}
\mathcal L \hat \rho=\kappa\left(2\hat a\hat \rho\hat a^{\dagger}-\hat a^{\dagger}\hat a\hat\rho-\hat\rho\hat a^{\dagger}\hat a\right)\,.
\end{equation}
The losses are assumed to be due to the mirror finite transmittivity, while spontaneous decay is neglected assuming that the atomic detuning exceeds the transition linewidth by several orders of magnitude. The Hamiltonian $\hat H$ is given by
 \begin{align}\label{Hinitial}
  \hat H=&\frac{\hat p^2}{2m}+\hat W_{\rm ext}(\hat x)+\hat H_{\rm opto}\,,
\end{align}
where the first term on right-hand side (RHS) is the kinetic energy, the potential $\hat W_{\rm ext}(\hat x)$ is periodic with period $\pi/k_0$ and tightly binds the atom at its minima,
$$\hat W_{\rm ext}(\hat x)=W_0\cos^2(k_0 \hat x)\,,$$
with $W_0$ the potential depth with the dimensions of an energy. Hamiltonian $\hat H_{\rm opto}$ includes the cavity degrees of freedom and their optomechanical coupling with the atomic motion, and reads
\cite{Domokos:2003,Schuetz:2013}
\begin{align}\label{Hopto}
\hat H_{\rm opto}=-\hbar \delta_c\hat a^{\dagger}\hat a+\hbar U_0\cos^2(\beta k_0\hat x)\hat a^{\dagger}\hat a+\hbar \zeta(\hat x) (\hat a^{\dagger}+\hat a)\,,
\end{align}
where frequency $U_0$ scales the dynamical Stark shift due to the coupling between atom and cavity mode, $U_0=g^2/\Delta_a$, with the vacuum Rabi frequency $g$, which determines the strength of the coupling between the dipole and one cavity photon. The frequency $U_0$ can be either positive or negative depending on the sign of $\Delta_a$. The last term on the RHS in Eq. \eqref{Hopto} corresponds to the effect induced by an external pump on the cavity mode. The pump, in particular, can couple either directly to the cavity, by impinging on a mirror, or via the atoms, which coherently scatter photons into the cavity mode. When the pump is set directly on the cavity mirror, the strength of this coupling is given by a constant value $\zeta(x)=\eta$. When instead the atom is transversally driven by the laser, then $\zeta(x)$ takes the form $$\zeta(x)=\cos(\beta k_0x)\Omega g/\Delta_a$$ and is thus weighted by the cavity spatial mode function at the atomic position. Moreover, it is proportional to the laser Rabi frequency $\Omega$, which determines the strength of the coupling between the dipole and the laser.

\subsubsection{Time-scale separation and effective dynamics}

We consider the limit in which there is a time-scale separation between cavity and atomic motion dynamics, and require that the inequality $|\kappa+{\rm i}\delta_c|\gg k\Delta p/m$ is fulfilled, where $\Delta p=\sqrt{\langle \hat p^2\rangle}$ is the variance of the atomic momentum, assuming that the cavity linewidth is much larger than the atom Doppler broadening \cite{Schuetz:2013}. We then identify the coarse-grained time scale $\delta t$, which is sufficiently short with respect to the time scale of the atomic external degrees of freedom and yet sufficiently long that during $\delta t$ the cavity field reaches a local steady state. 

The treatment is best illustrated in the Heisenberg picture and is detailed in Ref. \cite{Larson:2008,Habibian:2013}. We report here some relevant steps.
The equations of motion of the atom and of the cavity field operator, valid for the case in which the cavity is pumped by the laser, read
\begin{align}
\dot{\hat p}=&2\hbar kU_0\cos(k\hat x)\sin(k\hat x)\hat a^{\dagger}\hat a+2k_0W_0\cos(k_0\hat x)\sin(k_0\hat x)\label{pdot},\\
\dot{\hat a}=&-\kappa \hat a+{\rm i}(\delta_c-U_0\cos^2(k\hat x)) \hat a-{\rm i}\eta+\sqrt{2\kappa}\hat{a}_{\rm in}\label{a_with_Psi1}\,,
\end{align}
with $\hat{a}_{\rm in}(t)$ is the input noise operator, with $\langle \hat{a}_{\rm in}(t)\rangle=0$ and $\langle \hat{a}_{\rm in}(t)\hat{a}_{\rm in}^\dagger(t')\rangle=\delta(t-t')$ \cite{Walls}. Within the time step $\delta t$, with $\delta t\gg 1/|\delta_c+{\rm i}\kappa|$ but yet shorter than the atom's characteristic time scale, we identify the coarse-grained field operator $\hat{a}_{\rm st}$, which is defined by the equation
$$\int_t^{t+\delta t}\hat{a}(\tau)d\tau/\delta t\approx \hat{a}_{\rm st}\,,$$
such that  $\int_t^{t+\delta t}\dot{\hat a}_{\rm st}(\tau)d\tau=0$, with $\dot{\hat a}$ given in Eq. \eqref{a_with_Psi1}. The "stationary" cavity field is a function of the atomic operators at the same (coarse-grained) time, and in the limit where the quantum noise $\hat a_{\rm{in}}$ averaged over $\delta t$ can be neglected it takes the form
\begin{equation}\label{a_mf}
\hat{a}_{\rm st}\approx\frac{\zeta(\hat x)}{(\delta_c-U_0\cos^2(k\hat x))+{\rm i}\kappa}\,.
\end{equation}
A sufficient condition, for which this expression is correct, is that the mean intracavity photon number is larger than unity. A necessary condition, which originates from the statistical averaging at the basis of this treatment, is that $\kappa/\delta t \ll \zeta^2$. In this limit, the field at the cavity output reads
\begin{equation}
\label{a:out}
\hat{a}_{\rm out}=\sqrt{2\kappa}\hat{a}_{\rm st}-\bar{\hat a}_{\rm in}\,,
\end{equation}
and allows one to monitor the state of the atoms \cite{Walls,Mekhov:2007,Habibian:2013}. Using Eq.~\eqref{a_mf} for the field $\hat{a}$ in Eq.~\eqref{pdot} leads to an equation of motion for the atomic variables which depends solely on the atomic variables \cite{Larson:2008}. The corresponding effective Hamiltonian reads
\begin{align}
\label{H:eff}
 \hat H_{\rm eff}= \frac{\hat p^2}{2m}+W_0\cos^2(k_0\hat x)+  \hat V_{\text{eff}}(\hat x)\,,
\end{align}
where 
$$\hat V_{\text{eff}}(\hat x)=v_0 f(\hat x)\,.$$
Function $f(x)$ is given in Eq. \eqref{atan}, with now  $\delta_c'=\delta_c/\kappa$ and $C=U_0/\kappa$, thereby linking the parameters of our model to the microscopic theory. The energy $v_0$ takes a different form depending on whether the atom or the cavity is driven. When the latter is pumped, then
\begin{equation}
v_0=\frac{\hbar}{\kappa}\eta^2\,,
\end{equation}
while when the atom is transversally pumped it takes the form
\begin{equation}
v_0=\hbar\frac{\Omega^2}{\Delta_a}\delta_c' \,.
\end{equation}
The Hamiltonian \eqref{H:eff} is aperiodic, and contains the nonlinear coupling due to the cavity field in the functional form $f(x)$. It can be cast in a Hubbard form using the single particle Wannier basis $\{w_n\}$ of the external potential. Using this change of basis, in the tight-binding and single-band approximation, one obtains Eq. \eqref{Hfinal} from Eq. \eqref{H:eff}, where the onsite energy $\delta\epsilon_n$ is given by Eq. \eqref{Harper}, with $f(x)$ as defined in Eq. \eqref{atan}. The tunneling $t$ has the form
\begin{equation}
  t=\int_{-L_0/2}^{L_0/2} dx \ \ w_n(x)\left(\frac{-\hbar^2}{2m}\frac{d^2}{dx^2}+W_0\cos^2(k_0x)\right)w_{n+1}(x)\,.
 \end{equation}
We have verified that the site-dependent tunneling terms due to the cavity potential,
\begin{align}\label{t_sites}
  t_{n}=\int dx \ w_n(x) V_{\text{eff}}(x)w_{n+1}(x)\,,
\end{align}
are negligible for the parameters we choose and that we will specify in the next subsection. We remark that, while the resulting dynamics is coherent, its validity relies on a separation between the typical time scales of the cavity, which is intrinsically lossy, and the ones of the atomic motion. 

\section{Results}
\label{Sec:PD}

In this section we determine the phase diagram for the ground state of Hamiltonian \eqref{Hfinal}, analyze in detail the properties of the localization transition in the framework of CQED, and then discuss possible realizations with existing experimental setups of CQED with cold atoms. 

\subsection{Phase diagrams}

We determine the IPR, Eq. \eqref{IPR}, taking a lattice with open boundaries (hard walls) and choosing $\beta=\frac{\sqrt5-1}{2}$. The plots we show are evaluated for $L=233$. We checked that the IPR and the phase diagrams remain substantially unvaried when scaling up the lattice size $L$. For this system size, moreover, the behaviour of the Lyapunov exponent in the localized phase,  qualitatively reproduces the thermodynamic limit. We note here that, since the confining lattice has a minimum at $x=0$, after adding the perturbing potential of Eq. \eqref{atan} for $C<0$ the total potential exhibits a minimum at the center, while for $C>0$ the center is a local maximum (see Fig. \ref{Fig1}(b)). The symmetry by mirror reflection about the center, thus, gives that for $C<0$ the localized state is in the center, while for $C>0$ is a coherent superposition of two sites equally distant from $x=0$. In order to get an unique localized state for all values of $C$, for $C>0$ we take $f(x)=\arctan\left(-\delta_c'+C\sin^2(\pi \beta x/a)\right)$. This choice allows us to directly compare the localization transition for positive and negative values of $C$, thus to analyze the sole effect of the potential  minimum, which for $C>0$ is a narrow well while for $C<0$ is shallow about $x=0$.

\begin{figure}[htbp]
\begin{center}
  (a)\vtop{\vskip-0ex\hbox{\includegraphics[width=0.35\textwidth]{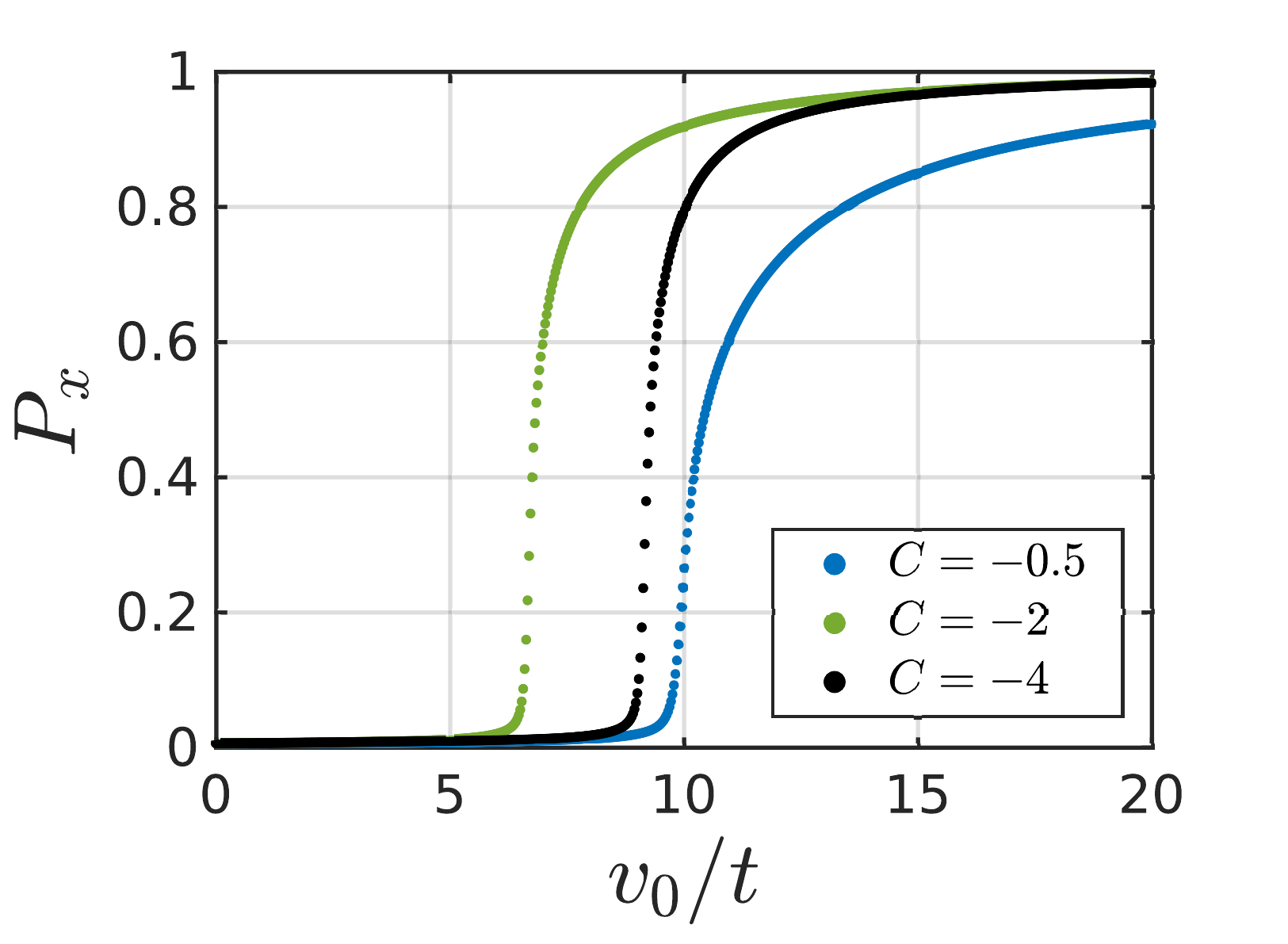}}}
  (b)\vtop{\vskip-0ex\hbox{\includegraphics[width=0.35\textwidth]{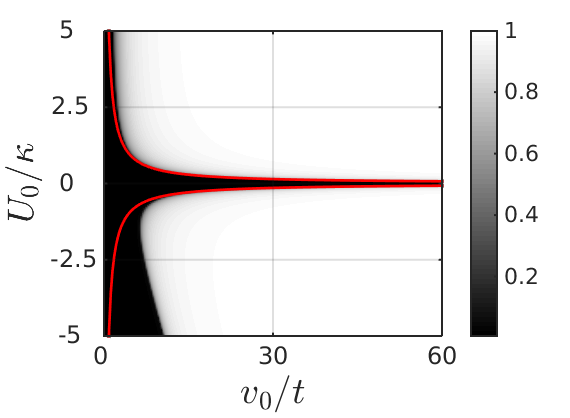}}}
\end{center}
\caption{(Color online) (a) Inverse participation ratio (IPR), Eq. (\ref{IPR}), as a function of $v_0$ (in units of $t$) for $\delta_c'=0$ and $C=-0.5,-2,-4$ (see legend). (b) Contour plot of the IPR as a function of  $v_0$ (in units of $t$) and of $C$, for $\delta_c'=0$. The solid lines (red) correspond to Eq. \eqref{v0:c}.\label{Fig:PD}}
\end{figure}

For all considered values of $C$ and $\delta_c'$ the functional behaviour of the IPR  as a function of $v_0$ exhibits a sharp transition, as visible in Fig. \ref{Fig:PD}(a) for various values of $C$. The critical value at which the transition occurs is given in good approximation by the one in Eq. \eqref{v0:c} for $|C|\ll 1$, while it differs from this value the larger $|C|$ becomes. This is clearly visible in Fig. \ref{Fig:PD}(b), which displays the contour plot of the IPR as a function of $v_0/t$ and $C$ for $\delta_c'=0$. Here, the solid lines correspond to Eq. \eqref{v0:c}, which predicts the transition value for the corresponding dual model, and is visibly shifted with respect to the transition we identify between extended (dark region) and localized state (light region).

\begin{figure}[htbp]
\begin{center}
  (a)\vtop{\vskip-0ex\hbox{\includegraphics[width=0.32\textwidth]{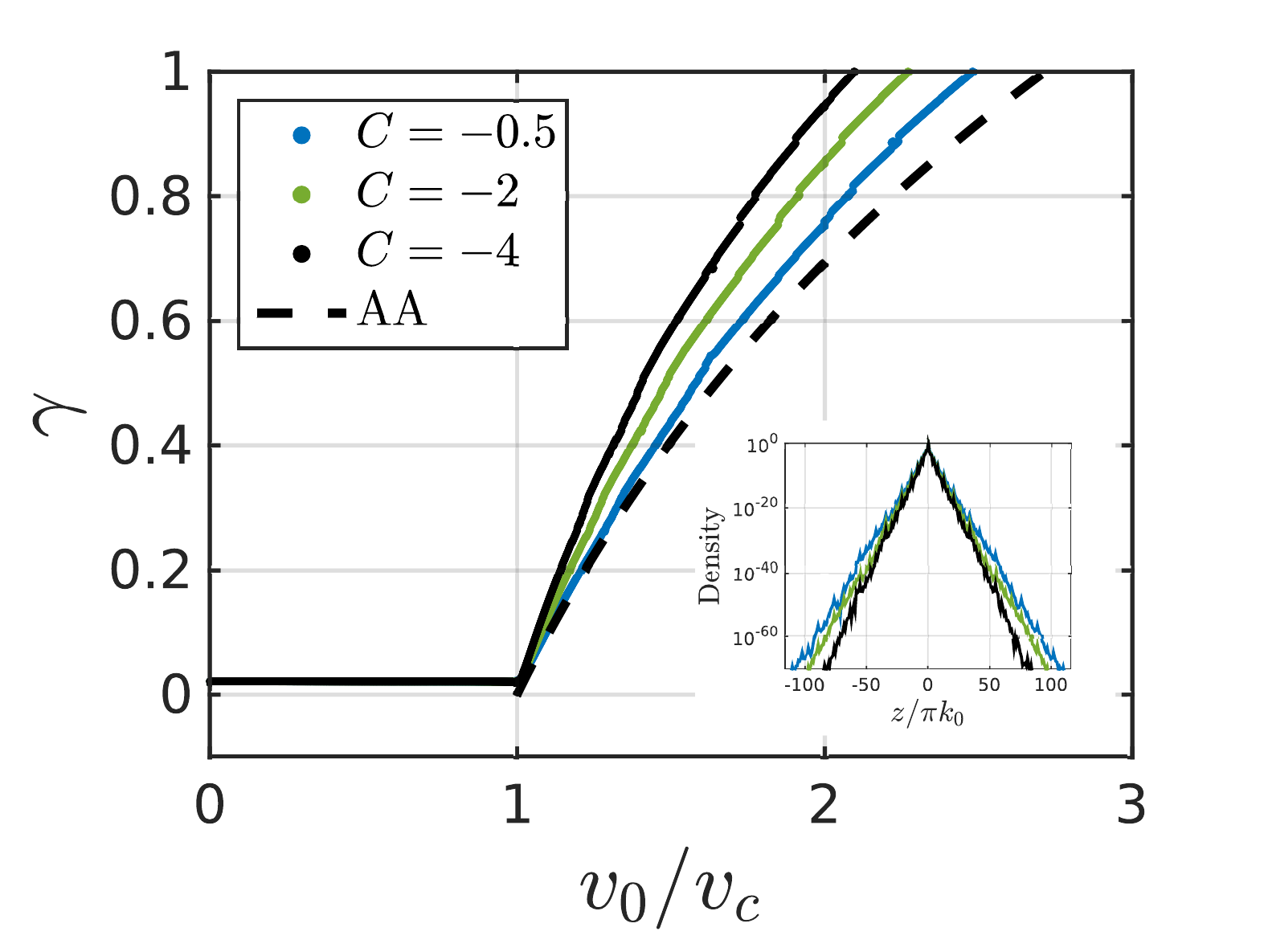}}}\\
  (b)\vtop{\vskip-0ex\hbox{\includegraphics[width=0.32\textwidth]{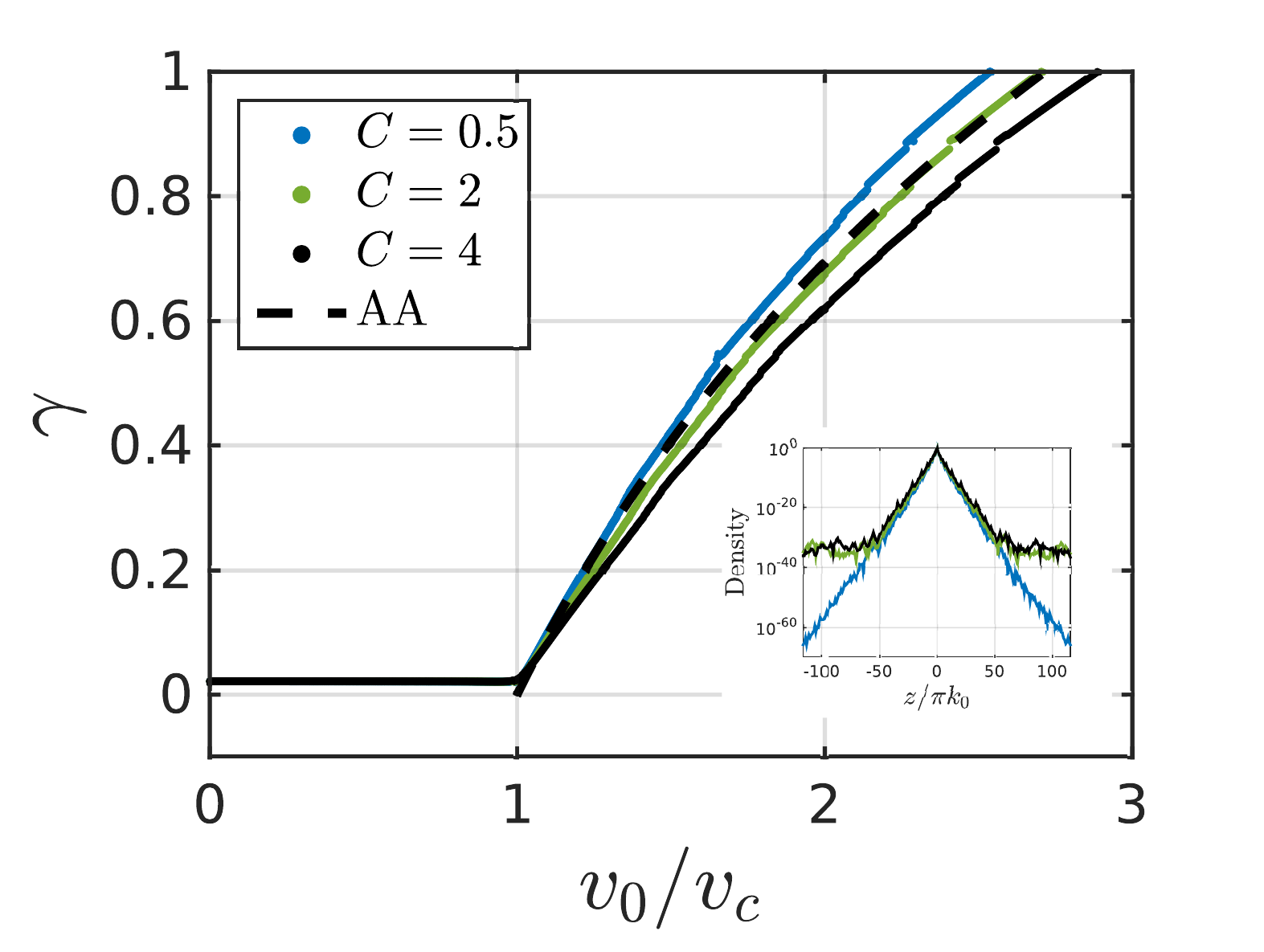}}}
  (c)\vtop{\vskip-0ex\hbox{\includegraphics[width=0.32\textwidth]{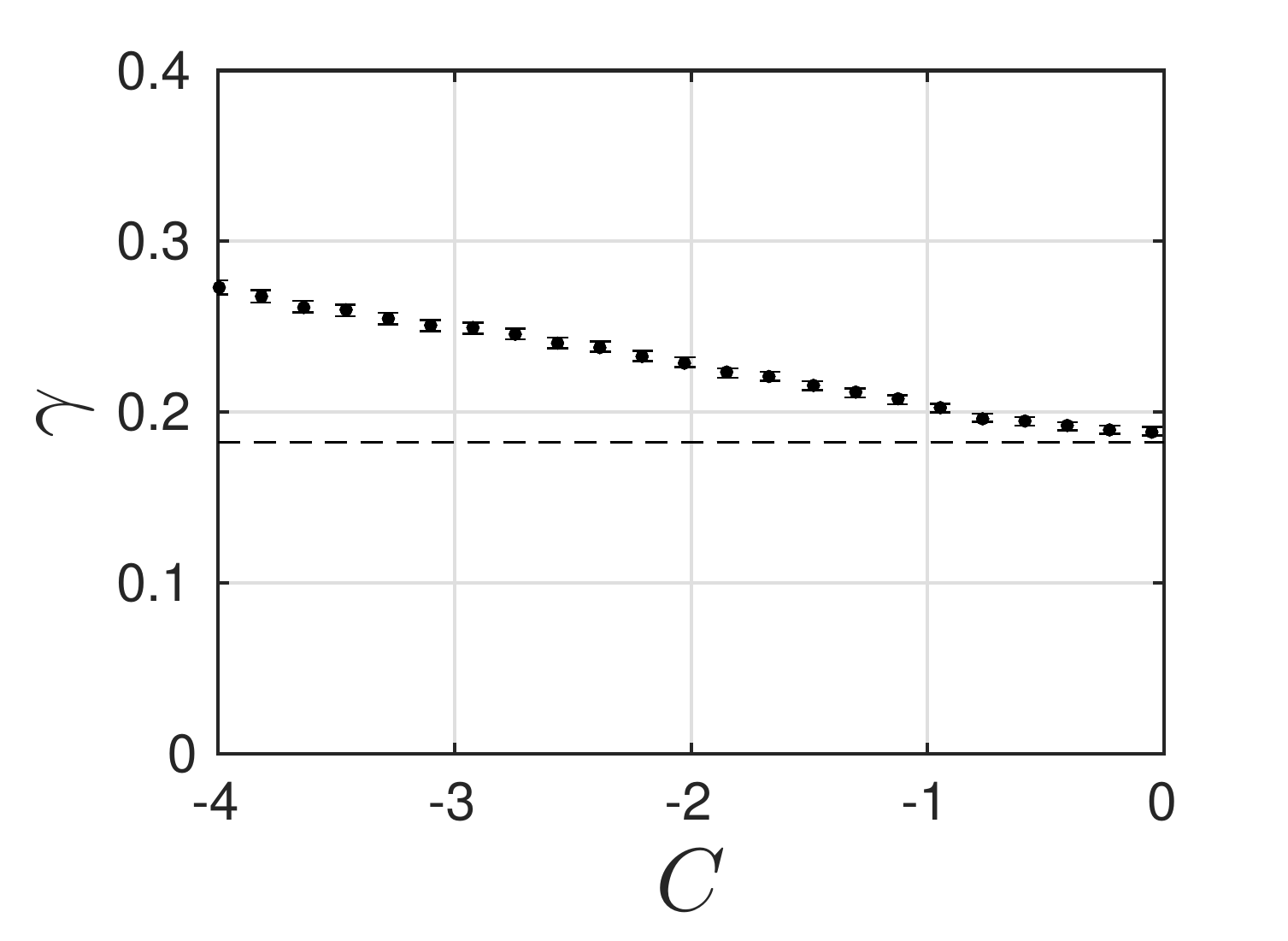}}}
  (d)\vtop{\vskip-0ex\hbox{\includegraphics[width=0.32\textwidth]{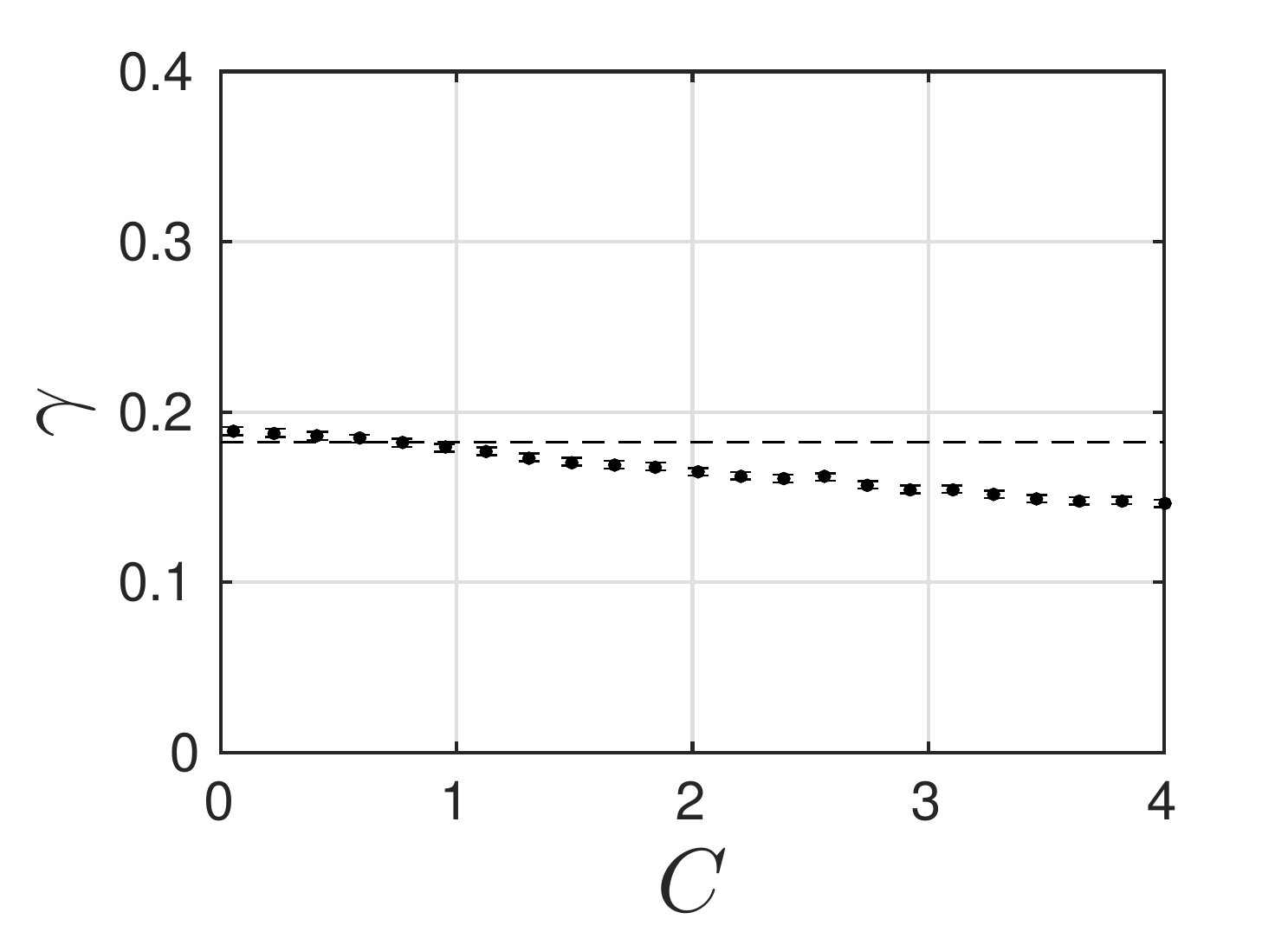}}}
\end{center}
\caption{(Color online) Lyapunov exponent as a function of $v_0$, in units of the critical depth $v_c$, which we extract from the numerical behaviour of the IPR in Fig. \ref{Fig:PD}, for $\delta_c'=0$ and for (a) $C=-0.5,-2,-4$ (b) $C=0.5,2,4$. The black dashed line corresponds to the functional behaviour of the Lyapunov exponent in Aubry-Andr\'e's model, Eq. \eqref{crit_exp}. The insets display the probability densities as a function of $x$ for the corresponding values of $C$ of the curves in the onset and for $v_0/v_c=2$. Subplots (c) and (d) displays the Lyapunov exponent as a function of $C$ for the fixed ratio $v_0/v_c=1.2$, the horizontal line indicates the value predicted by Eq. \eqref{crit_exp} ($v_c$ depends on $C$, for each value of $C$ it is extracted from the curves of the IPR as in Fig. \ref{Fig:PD}(a)). \label{IPR_dc0u0n}}
\end{figure}

We analyze the properties at the transition by plotting the probability density as a function of $x$ and observe that in the localized phase it always exhibits an exponential decay, although for $|C|>1$ we also find that for same parameter regimes at large distances the density profile shows an extended component (see inset of Fig. \ref{IPR_dc0u0n} (b) and Fig. \ref{Lyap_dc-2}). We have checked that this uniform background is not a numerical artifact. Typical probability densities are shown in the insets of Fig. \ref{IPR_dc0u0n}(a) and (b). We remark that deviations from a purely exponential profile have been observed in the localized phase of a Bose-Einstein condensate of weakly interacting atoms, where the ground state was the superposition of several localized states \cite{Lewenstein:BEC} due to the effect of interactions. In our case, the observed density profile can be viewed as the overlap between a localized and an extended state. This behaviour is due to the higher harmonics of the cavity potential, Eq. \eqref{atan}: Indeed, we checked that the background appears already by truncating the Taylor expansion of Eq. \eqref{atan} in $|C|$ to order. 

Figure \ref{IPR_dc0u0n}(a) and (b) display the Lyapunov exponent $\gamma$ as a function of $v_0$ for $C<0$ and $C>0$, respectively. The values are extracted by performing a fit of the central localized region of the density profiles (see insets). This procedure introduces  an uncertainty in the determination of the Lyapunov exponent, which is not shown here since it is comparable with the size of the markers. The dependence of 
$\gamma$ on $C$ for fixed $v_0/v_c$ is shown in subplots (c) and (d), where now the error bars give the uncertainty in the value we fitted. For $C<0$ the Lyapunov exponent (and thus localization) increases with $|C|$ and is larger than the value of Eq. \eqref{crit_exp}, to which it tends for $C\to 0^-$.The behaviour is qualitatively different for $C>0$, as visible in subplot (d):  As $C$ is increased from 0, the Lyapunov exponent decreases monotonically from the value of Aubry's model. The curve seems to tend to a nonvanishing asymptotic constant value for $C\to\infty$, which is the limit of a sequence of infinitely narrow wells as shown in Fig. \ref{Fig1} (b). 

\begin{figure}[htbp]
\begin{center}
  \includegraphics[width=0.35\textwidth]{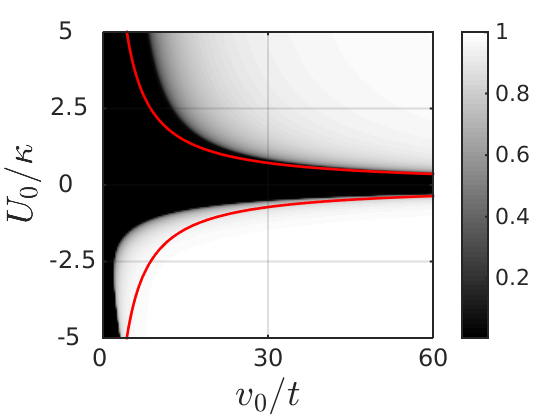}
\end{center}
\caption{(Color online) Contour plot of the IPR as a function of  $v_0$ (in units of $t$) and of $C$, for $\delta_c'=-2$. The solid lines (red) correspond to Eq. \eqref{v0:c}.\label{IPR_dc-2}}
\end{figure}

\begin{figure}[htbp]
\begin{center}
  (a)\vtop{\vskip-0ex\hbox{\includegraphics[width=0.35\textwidth]{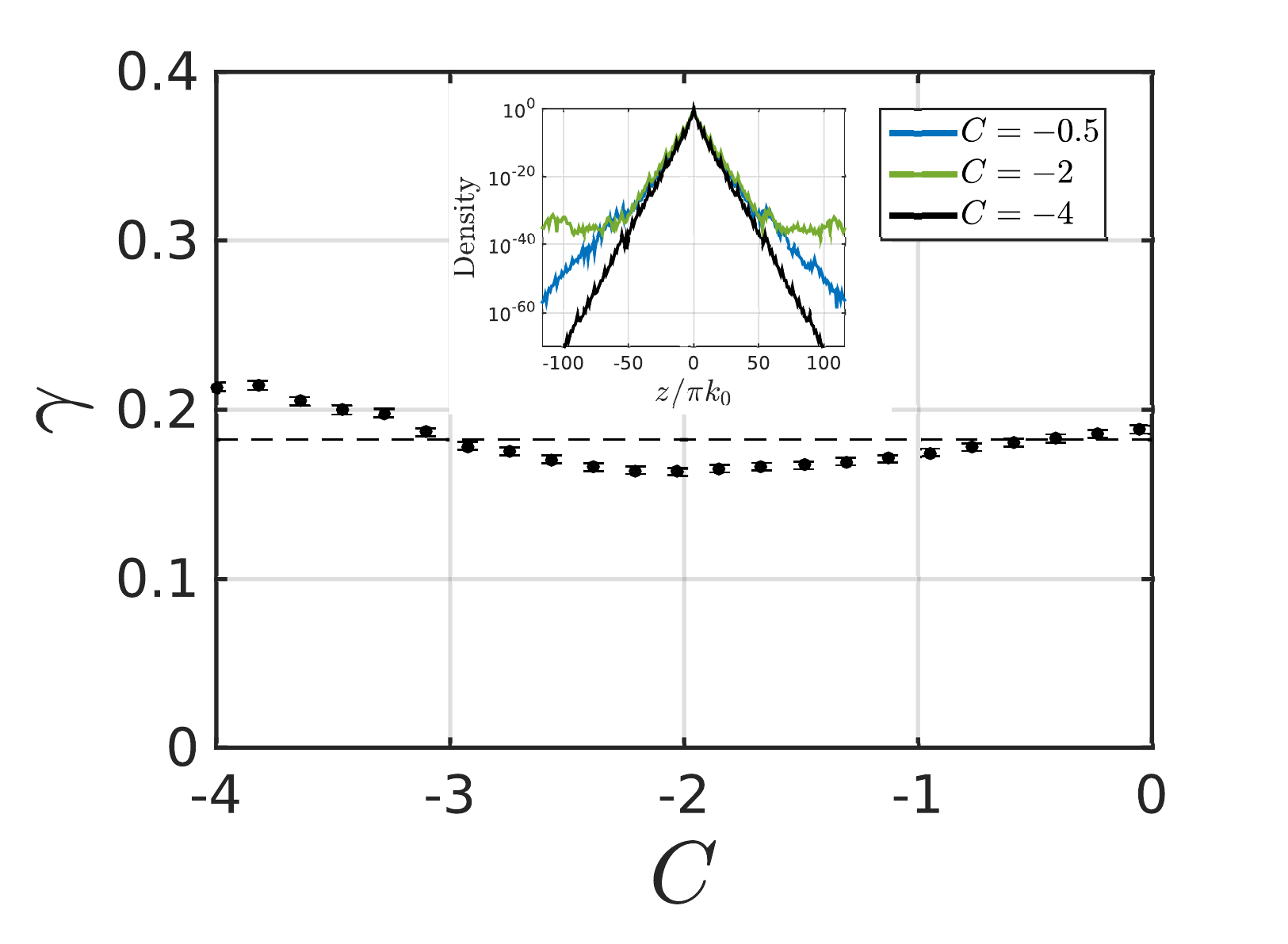}}}\\
  (b)\vtop{\vskip-0ex\hbox{\includegraphics[width=0.35\textwidth]{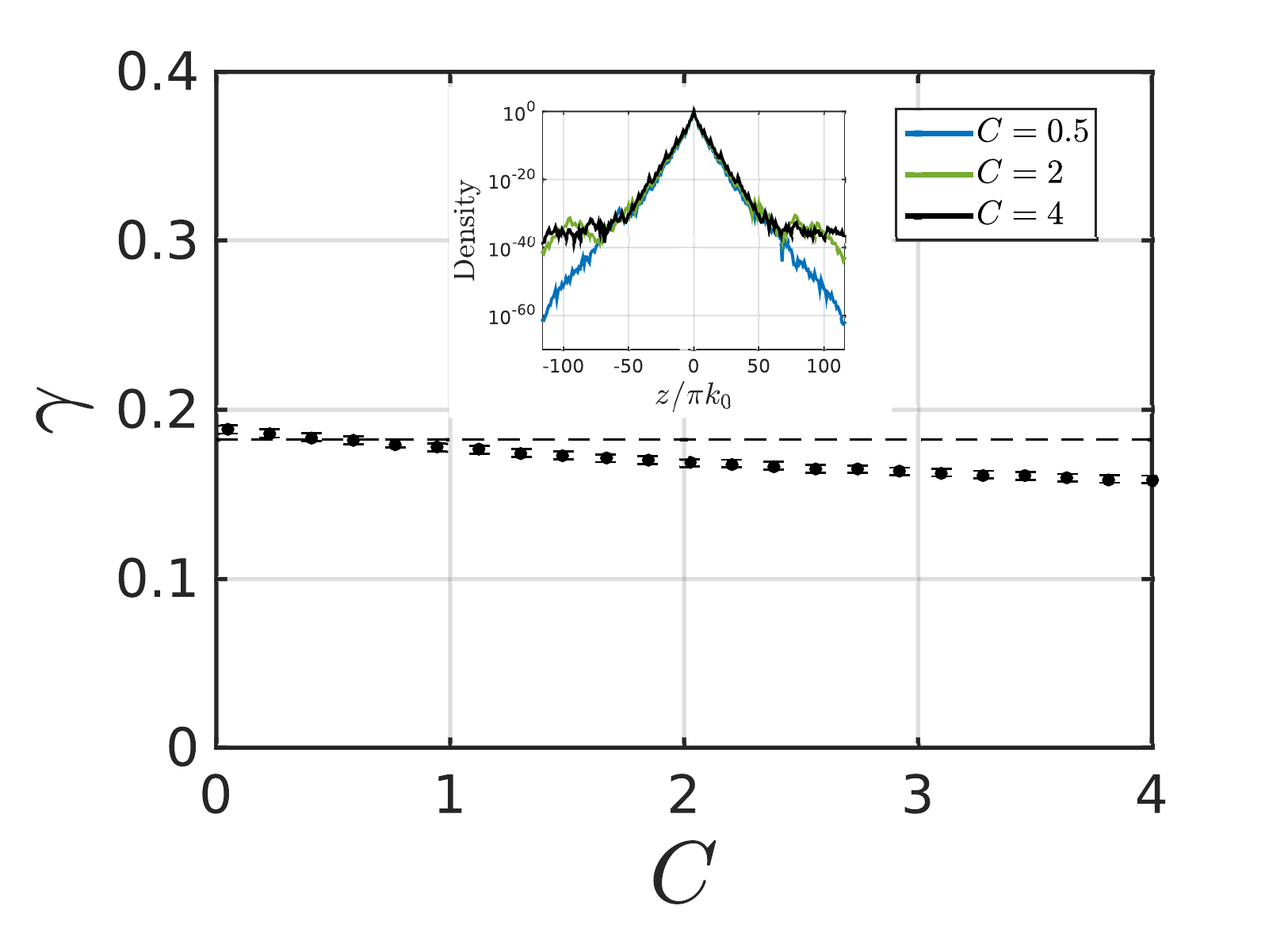}}}
\end{center}
\caption{(Color online) Lyapunov exponents as a function of $C$ for $v_0/v_c=1.2$, the black horizontal dashed line indicates the value predicted by Eq. \eqref{crit_exp}. The insets display probability densities as a function of $x$ for different values of $C$ and for the fixed ratio $v_0/v_c=2$.\label{Lyap_dc-2}}
\end{figure}

We now explore the dependence of $\gamma$ and of the IPR on the detuning $\delta_c'$. We have checked several values and take $\delta_c'=-2$ in order to provide a representative example. For this value we analyze the IPR (Fig. \ref{IPR_dc-2}) and the corresponding dependence of the Lyapunov exponents on $C$ (Fig. \ref{Lyap_dc-2}). 
The contour plot shows that for $C<0$ the extended phase shrinks with respect to the case $\delta_c'=0$ (Fig. \ref{Fig:PD}(b)), the smaller critical value $v_c$ is found at about $C\sim -2$. Correspondingly, the Lyapunov exponent as a function of $C$ possesses a minimum at the same value of the cooperativity. 
This value is given by a root of the function $f(x)$, Eq. \eqref{atan}, for $\cos^2(kx)\approx 1$, which is fulfilled when  the atom is localized at the minimum of the total potential. This root is a resonance which maximizes the intracavity photon number when the atom is in a localized state, as we will show below. 

\subsection{Experimental realization}

Single atoms and ions have been trapped inside cavities and cooled to very low temperatures  \cite{Walther:2006,Kimble:2005}, the dispersive coupling with the cavity field as in Eq. \eqref{H:eff} has been realized \cite{Ritsch:2013}. These implementations rely on the existence of an external trapping potential, that is typically harmonic. This breaks the discrete translational invariance along the direction of motion thus drastically changing the properties of the extended state. However, a sufficiently shallow trap does not affect the transition to localization as long as the size of the localized state is much smaller than the harmonic oscillator length \cite{Roati2008}. Inclusion of the harmonic confinement would be a straightforward extension of the present model. We do not include the harmonic trapping in the present work since under typical experimental conditions (see eg   Ref.\cite{Wolke:2012}) the harmonic oscillator length ($\sim0.6$mm for a trapping frequency of $\omega_{\rm{ho}}=2\pi\times25$kHz of $^{87}$Rb atoms) is much larger than the size of the localized wavefunction ($\sim 2\mu$m for $\gamma=0.2$ and $k_0=2\pi/830$nm). 

The transition to localization with cold atoms can be revealed by means of time-of-flight measurement, as realized in Ref. \cite{Roati2008}, or in-situ imaging \cite{Sherson2010,Haller2015}. Another possibility is to analyze the spectrum of light emitted by the resonator, since this contains the information about the system excitations and allows one to monitor the dynamics \cite{Cormick:2013}. 

The conditions for the time scale separation we performed in Sec. \ref{Sec:CQED} are fullfilled, provided the shot noise component of the cavity field can be neglected over the time scale of the motion, which leads to a condition on the number of intracavity photon and on the atom's kinetic energy. Figures \ref{PD_cav}(a) and (b) display the phase diagram obtained from the IPR, here reported as a function of the pump strength $\eta$, of the frequency $U_0$ and of the detuning $\delta_c$, for the  parameters of the setup of Ref. \cite{Wolke:2012,Klinder:2015}. The transition to localization can be observed for cooperativity $|C|=|U_0|/\kappa\simeq 1$, and thus require the strong coupling at the single atom level. We further determine the corresponding mean intracavity photon number $\bar n$, according to the equation
\begin{align}\label{our_ncav}
  \bar n&=\langle\Psi_0|\hat a^{\dagger}_{\text{st}}\hat a_{\text{st}}\ket{\Psi_0}\nonumber\\ &\simeq\sum_{m=1}^{L} |\langle m|\Psi_0\rangle|^2\int dz \ w_m(z)^2 \frac{\zeta^2}{\left(\delta_c-U_0\cos^2(k_cz)\right)^2+\kappa^2}\,.
\end{align}
Its form shows that the root of Eq. \eqref{atan} is an optomechanical resonance in the cavity field \cite{StamperKurn}.  

\begin{widetext}
Figures \ref{PD_cav}(c) and (d) show the intracavity photon number for the parameters of the phase diagrams in subplots (a) and (b), respectively. 
\begin{figure}[h]
\begin{center}
  (a)\vtop{\vskip-0ex\hbox{\includegraphics[width=0.32\textwidth]{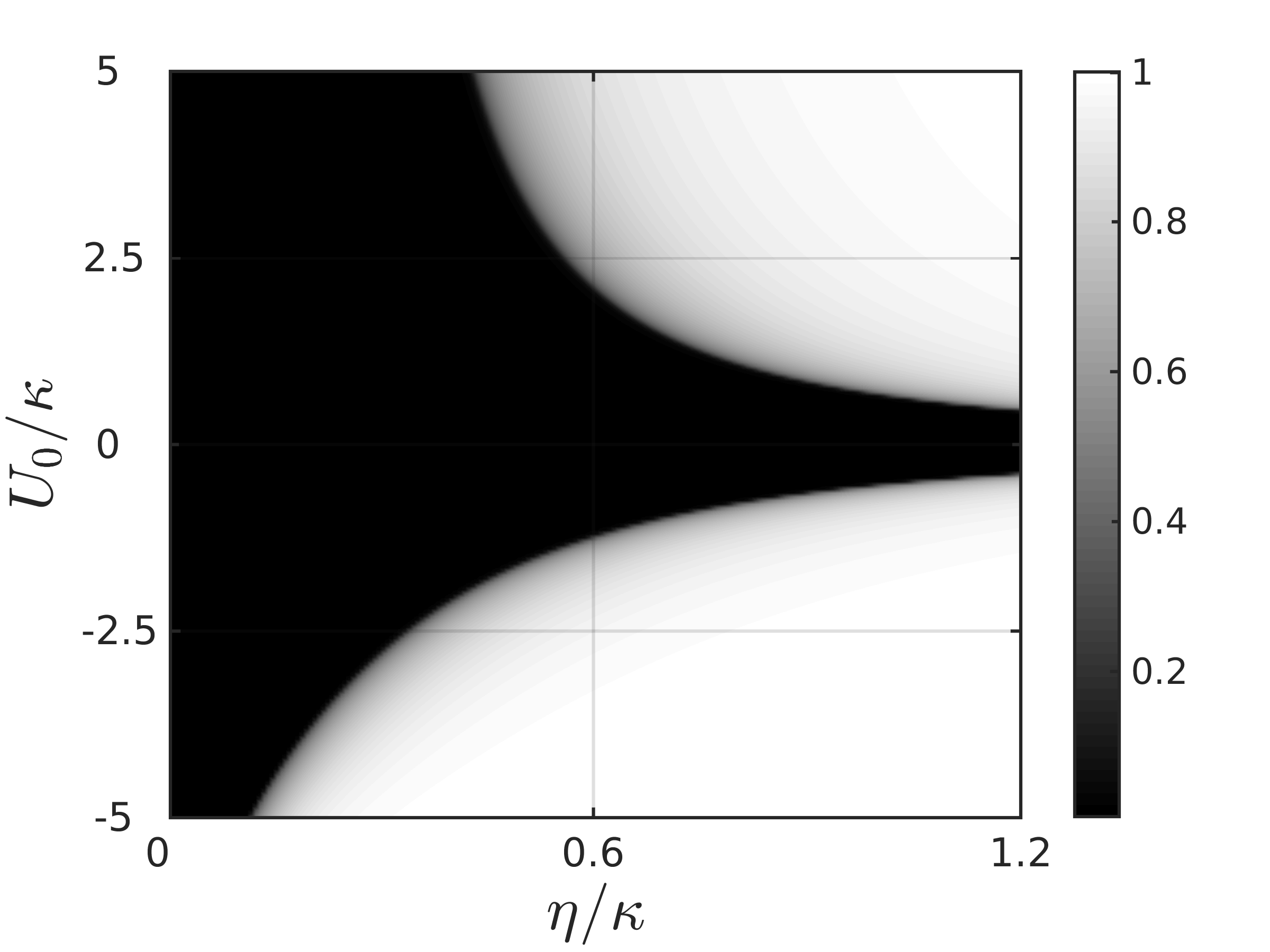}}}
  (b)\vtop{\vskip-0ex\hbox{\includegraphics[width=0.32\textwidth]{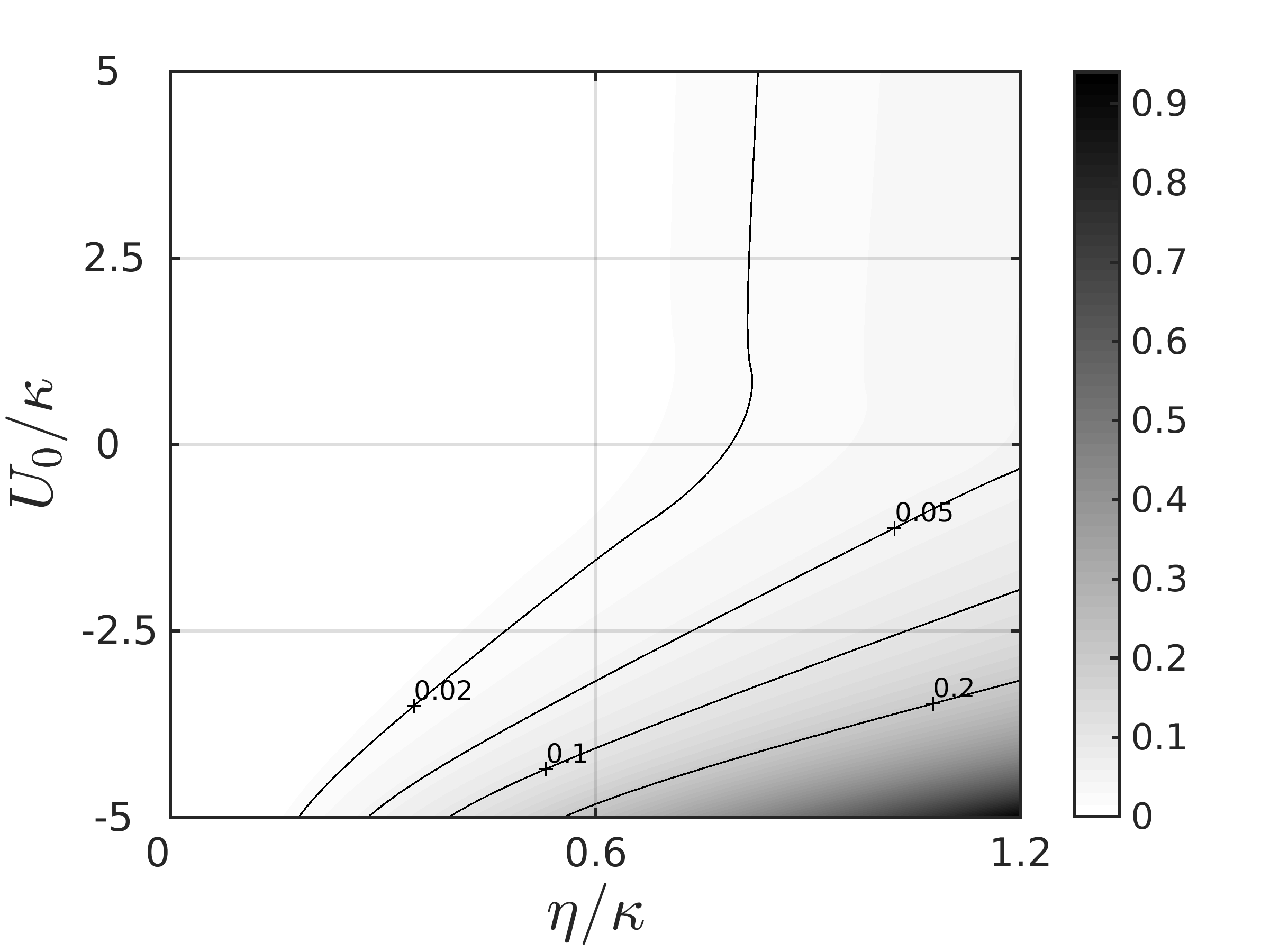}}}\\
  (c)\vtop{\vskip-0ex\hbox{\includegraphics[width=0.32\textwidth]{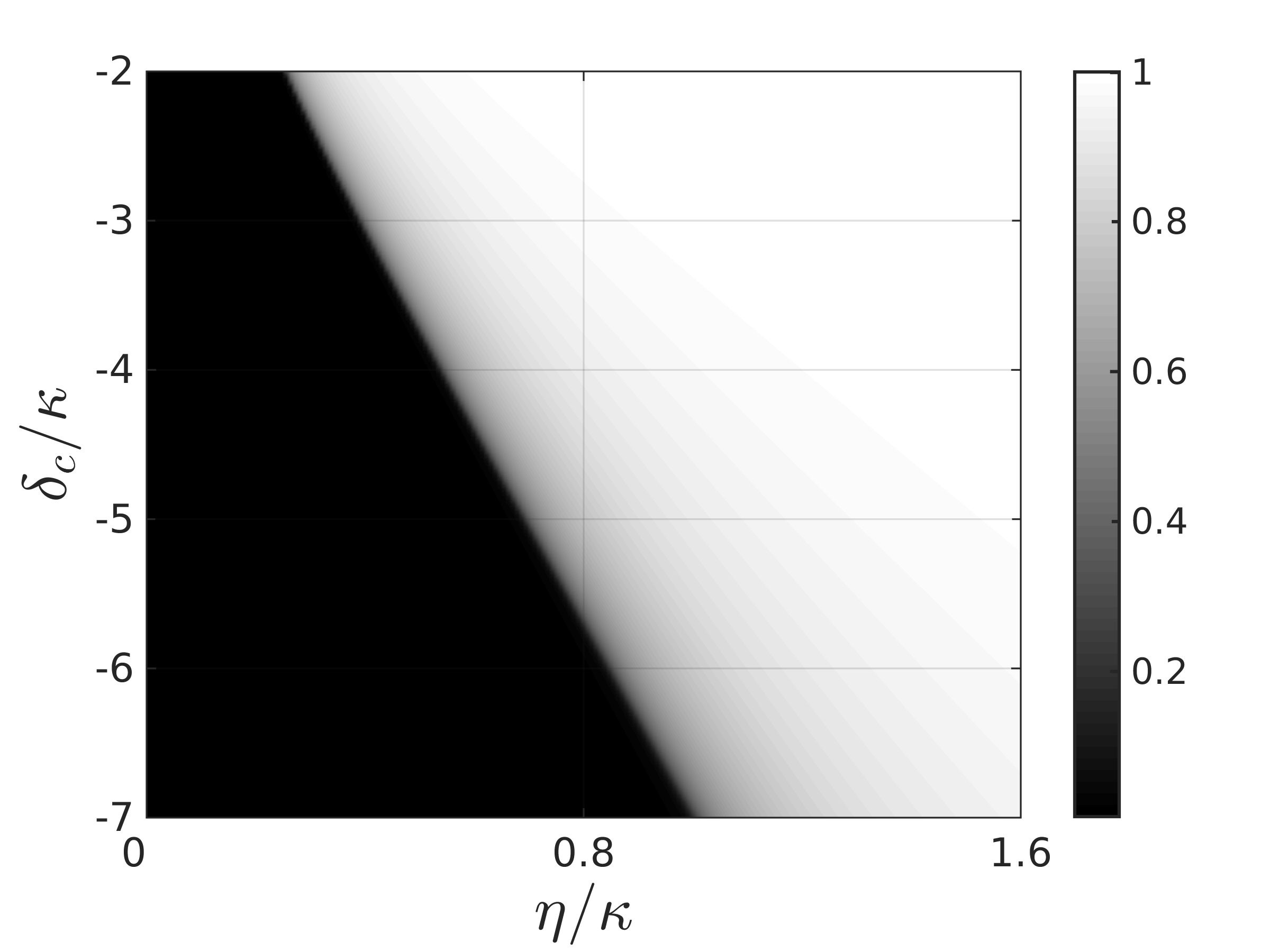}}}
  (d)\vtop{\vskip-0ex\hbox{\includegraphics[width=0.32\textwidth]{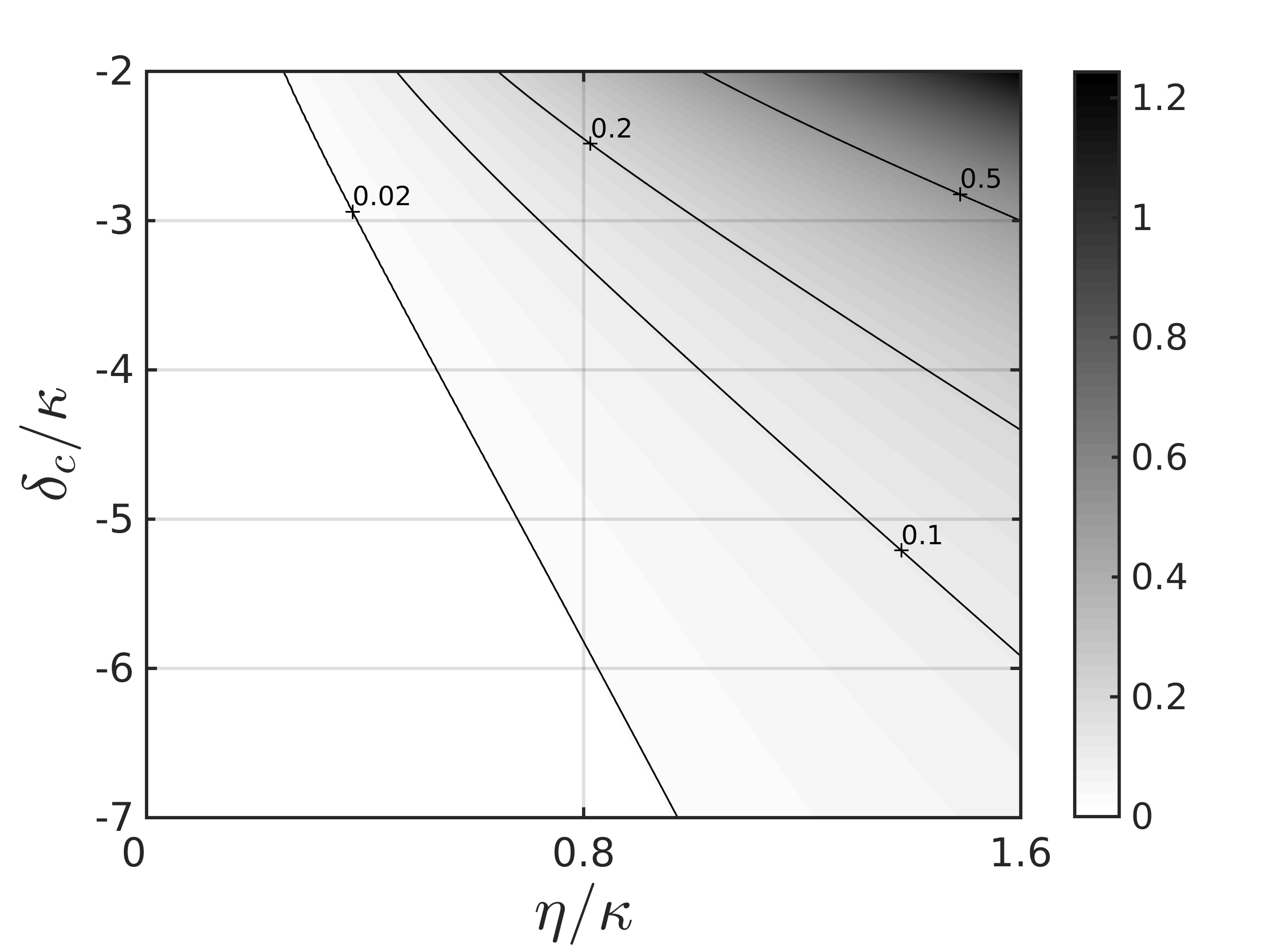}}}
\end{center}
\caption{(a) Inverse participation ratio, Eq. \eqref{IPR}, and (b) mean intra cavity photon number $\bar n$, Eq. \eqref{our_ncav}, as a function of the parameters $\eta$ and $U_0$ (in units of $\kappa$) in the setup where the resonator is driven and for detuning $\delta_c=-5.5\kappa$. Here, $\eta$ is the strength of the laser and $U_0$ is the strength of the optomechanical coupling. In (a) and (b) the potential depth is fixed to $V_0=-15E_r$, where $E_r$ is the recoil energy associated with the $D$ line of $^{87}$Rb atoms. Subplots (c) and (d) show the IPR and $\bar n$ as a function of $\eta$ and $\delta_c$ (in units of $\kappa$) for $U_0/\kappa=-1$ and $V_0=-14E_r$. The other parameters are the number of sites $L=233$ and $\beta=\frac{\sqrt5-1}{2}$. For the parameters of Ref. \cite{Wolke:2012,Klinder:2015}, where $\kappa\approx E_r/\hbar$, the time-scale separation at the basis of our model is warranted when the detuning, $|\delta_c|>E_r/\hbar$ and the atoms temperature, $T< 1\mu$K. \label{PD_cav}}
\end{figure}
The signal to noise ratio can be increased by confining $N$ bosonic atoms in the resonator since the total number of photons scales linearly with the number of atoms $N$. The dynamics we predict would scale up as long as the atoms form an ideal Bose gas, under similar conditions as the ones realized in the LENS experiment \cite{Roati2008, Modugno2009}, such that the onsite interaction is much smaller than the average kinetic energy at the localization transition. We further remark that the cavity nonlinearity also scales linearly with $N$, so that sufficiently large atomic samples allow one to reach the strong coupling regime. In this case, however, the cavity mediates an effective interaction between the atoms \cite{Larson:2008}. In absence of other type of interactions, such as for instance at a Feshbach resonance, one may expect that the ground state will be very close to the single-atom ground state, with a modified excitation spectrum.
\end{widetext}

\section{Conclusions}
\label{Sec:Conclusions}

We have analyzed the localization transition in a modified Aubry-Andr\'e model, where the secondary potential, whose periodicity is incommensurate with the confining lattice, is due to the coupling with a high finesse resonator. Its effective optomechanical potential consists of a trascendental function of the atomic position, which results from the sum of all the harmonics when the light scattered by the atom backacts on the atomic position. In this limit, the localization we predict is self-induced by the atom. We find that it preserves several features of the Aubry-Andr\'e model. Novel features are the shift of the localization in the phase diagram and the behaviour of the Lyapunov exponent, which is a function of the cooperativity and shows peculiar features close to the parameters where the system exhibits optomechanical resonances. 

The localization-delocalization transition we predict can be measured with ideal bosonic gases confined in resonators for existing setups \cite{Wolke:2012,Klinder:2015}. Our study sheds light into the effect of nonlinearities in the quantum regime and complements the studies on glassiness \cite{Habibian:2013} and static friction \cite{Fogarty:2015} in interacting gases induced by cavity backaction in frustrated geometries. 

\begin{acknowledgments}
The authors acknowledge discussions with Stefan Sch\"utz, Simon J\"ager, and Astrid Niederle, and financial support by the German Research Foundation (DFG, Quantum crystals of matter and light"), the German Ministry of Education and Research (BMBF "Q.com"), the Laboratoire d'Alliances Nanosciences-Energies du futur (LANEF) and the French-German University (UFA/DFH). AM ackowledges financial support from the ANR projects Mathostaq (ANR-13-JS01-0005-01) and  SuperRing (ANR-15-CE30-0012-02).
\end{acknowledgments}

\end{document}